Article
# EPR Correlations Using Quaternion Spin

Bryan Sanctuary †

Chemistry Department, McGill University, Montreal, QC H3A 0G4, Canada; bryan.sanctuary@mcgill.ca
† Retired professor.

**Abstract:** We present a statistical simulation replicating the correlation observed in EPR coincidence experiments without needing non-local connectivity. We define spin coherence as a spin attribute that complements polarization by being anti-symmetric and generating helicity. Point particle spin becomes structured with two orthogonal magnetic moments, each with a spin of $\frac{1}{2}$—these moments couple in free flight to create a spin-1 boson. Depending on its orientation in the field, when it encounters a filter, it either decouples into two independent fermion spins of $\frac{1}{2}$, or it remains a boson and precedes without decoupling. The only variable in this study is the angle that orients a spin on the Bloch sphere, first identified in the 1920s. There are no hidden variables. The new features introduced in this work result from changing the spin symmetry from SU(2) to the quaternion group, $Q_8$, which complexifies the Dirac field. The transition from a free-flight boson to a measured fermion causes the observed violation of Bell's Inequalities and resolves the EPR paradox.

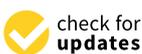

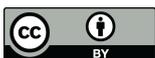

## 1. Introduction

We consider the correlation obtained from coincidence EPR experiments [1–4]. Here, we define an EPR pair [5] as two particles initially in a singlet state that separate [6]. In that process, we assume no non-local entanglement persists, so the pair forms a product state.

We have developed a spin theory [7], which includes a Pauli bivector, $i\sigma$ [8]. From this change, spin becomes structured and geometrically equivalent to a photon, Figure 1 [7]. It has two orthogonal spins $\frac{1}{2}$ axes, $\Sigma_3$ and $\Sigma_1$, which are reflections of each other, and in free-flight, these couple to give a composite boson spin of magnitude 1, $\Sigma_{13}$. We show that the observed violation of Bell's Inequalities (BI) [9] results from the decoupling of this boson spin-1 in a polarizing field, giving the usual fermion spin-$\frac{1}{2}$. The process epitomizes and formalizes the wave-particle duality for spin.

We refer to the two-state point particle spin-$\frac{1}{2}$ as a measured electron, $e_F^-$, a fermion, which is the solution to the Dirac equation [10]. It has two states of up and down, $|\pm, \hat{\mathbf{n}}\rangle$ where $\hat{\mathbf{n}}$ is a vector on the Bloch sphere. The complementary property to the measured spin polarization is the helicity [8]. The boson electron, $e_B^-$, exists only in free flight (isotropy) and displays helicity, which spins the axis of linear momentum either L or R. As a boson electron encounters a polarizing field, it transitions to the fermion electron.

This work is motivated by rejecting as nonphysical the notion of non-locality as asserted by Bell's Theorem [11]. This logically means that a local rational alternative to non-locality must exist. Here, Q-spin [7], which is on the same mathematical basis as the Dirac approach, accounts for the violation of Bell's Inequalities (BI) [9] without non-locality, thereby disproving Bell's Theorem by counterexample. Bell's Theorem underpins the field of Quantum Information theory, which accepts non-local entanglement as a property of Nature, see e.g., [12]. This has spawned major research efforts to exploit entanglement [13] and quantum teleportation [14]. For example, the quantum internet is presently at the theoretical stage but is primarily concerned with internet security using Quantum Key

Distribution [15–17]. Since quantum teleportation [14] rests entirely on Bell's Theorem, then this and all research based upon quantum teleportation requires reassessment.

These changes result from complexifying the Dirac field, leading to non-Hermitian operators essential to the formulation of Q-spin. Non-Hermitian states are not new. An extensive review is given by [18], and their use in optics is given by [19].

After discussing the origin of EPR correlation, we review quaternion, or Q-spin [7], without details. We then use Q-spin to simulate the EPR correlation, which reproduces the violation of BI without non-locality. The computer simulation generates the correlation from polarization and coherence, giving a CHSH correlation [1] of respectively 2 and 1 for a total of 3. Using a singlet state, QM gives CHSH = $2\sqrt{2} = 2.828$. The simulation treats all EPR pairs, with no missing coincidences, and they are statistically independent.

This paper is the third of four in which quaternion spin, or Q-spin, is presented. In the first paper [7], the Dirac equation is modified to include a bivector, $i\sigma_2 = \sigma_3 \sigma_1$, which is the origin of the complementary property to spin polarization, the helicity. The second paper [8] shows that the correlation between an EPR pair is conserved after separation. The fourth [20] summarizes some consequences of using Q-spin, which replaces Dirac's two point-particles, the matter–antimatter pair, with one structured particle. Q-spin is one particle with four states, whereas Dirac's equation gives two particles with two states each.

**A spin $\frac{1}{2}$ in free flight.**

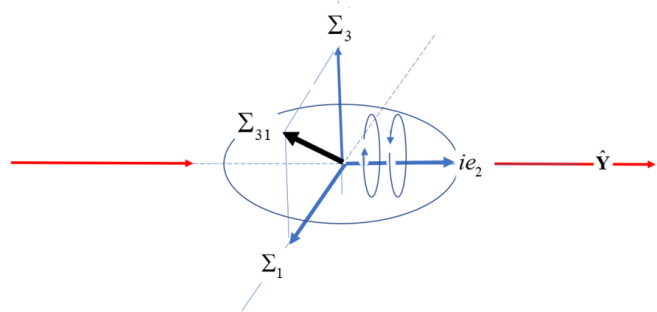

**Figure 1.** Two properties of spin: its polarization perpendicular to its helicity. The helicity is in the direction of propagation and averages out the polarizations.

## 2. EPR Correlation

Figure 2 plots the correlation arising from an EPR pair versus $\theta_{ab} = (\theta_a - \theta_b)$, where the angles are the filter settings at Alice and Bob. The coincidence data give a cosine similarity, which violates BI. In contrast, a product state does not violate BI with a CHSH = 2. The figure displays this as a triangle. Subtracting the two separates the part responsible for the BI violation, with CHSH = 0.828. We call this the mustache function, and Bell's Theorem [11], asserts that this can only arise from non-locality. In contrast, here, we attribute the missing correlation to quantum coherence, which until this work had not been formulated. This is easily seen from the usual singlet state [21],

$$|\Psi_{12}\rangle = \frac{1}{\sqrt{2}}[|+\rangle_1|-\rangle_2 - |-\rangle_1|+\rangle_2] \quad (1)$$

and forming the outer product of Equation (1) to give a $4 \times 4$ matrix and defines the pure state operator, $\rho_{12}$,

$$\rho_{12} = |\Psi_{12}\rangle\langle\Psi_{12}| = \frac{1}{2}\begin{pmatrix} 0 & 0 & 0 & 0 \\ 0 & 1 & -1 & 0 \\ 0 & -1 & 1 & 0 \\ 0 & 0 & 0 & 0 \end{pmatrix} \quad (2)$$

The off-diagonal components, $|\pm\rangle\langle\mp|$, are responsible for the mustache function. Using Equation (2), we obtain the EPR correlation from an entangled state as shown in [7],

$E(a, b) = \mathbf{a} \cdot \langle \sigma^1 \sigma^2 \rangle \cdot \mathbf{b} = -\cos \theta_{ab}$. Dropping the two off-diagonal terms gives a product state, $E(a, b) = \mathbf{a} \cdot \langle \sigma^1 \rangle \langle \sigma^2 \rangle \cdot \mathbf{b} = -\cos \theta_a \cos \theta_b$. The unit vectors, **a** and **b** denote the filter vectors.

In EPR coincidence experiments, the observed correlation is inconsistent with the product state but gives the complete correlation. One of the many statements of the EPR paradox [22] is that entanglement must exist over spacetime. Bell's Theorem justifies this [11], but how such non-local connectivity is maintained is not understood and defies rational explanation [23–25]. Introducing the bivector overcomes these difficulties, which Geometric Algebra justifies [26].

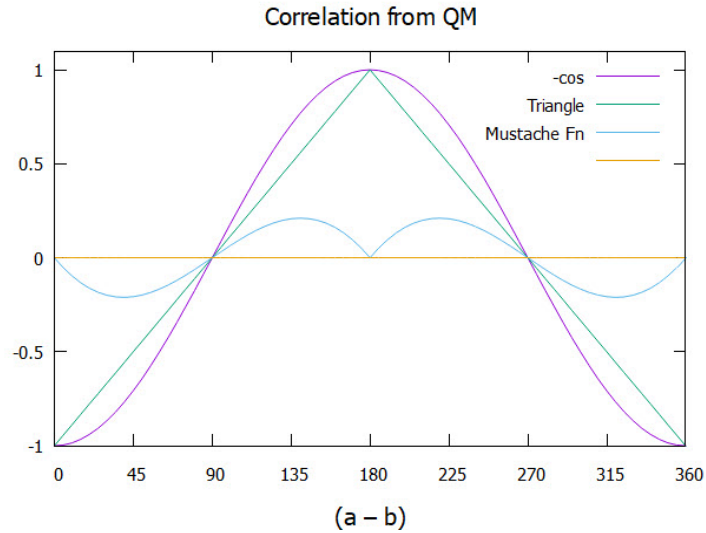

**Figure 2.** The full correlation from quantum mechanics with CHSH = $2\sqrt{2}$ = 2.828 is decomposed into product states with CHSH = 2, and the difference between the two gives the mustache function with CHSH = 0.828.

The geometric product is the sum of the symmetric scalar product and the anti-symmetric wedge product. This leads to the well-known relationship between the spin components of

$$\sigma_i \sigma_j = \delta_{ij} + \varepsilon_{ijk} i\sigma_k. \quad (3)$$

The bivector, $i\sigma_k$, plays no role in determining the observed spin polarization because it traces out. Define the helicity [8] as the complementary attribute to the polarization,

$$\underline{\underline{\mathbf{h}}} = \underline{\underline{\varepsilon}} \cdot i\sigma \quad (4)$$

Using this anti-symmetric, anti-Hermitian second-rank tensor operator in the same calculation as given above for a product state gives the missing correlation [8],

$$\mathbf{a} \cdot \left\langle \underline{\underline{\mathbf{h}}}^1 \right\rangle \cdot \left\langle \underline{\underline{\mathbf{h}}}^2 \right\rangle \cdot \mathbf{b} = -\sin \theta_a \sin \theta_b \quad (5)$$

Adding this to the product of cosines recovers the full singlet correlation, $-\cos \theta_{ab}$, suggesting the observed violation of BI results from helicity. This is an example of the Conservation of Geometric Correlation [8]. Non-locality plays no role here.

*2.1. Quaternion Spin*

The details of this section are found in [7].

The Dirac equation formulates spin as a four-dimensional real field represented by the four anti-commuting gamma matrices, $\gamma^\mu$. Dirac found two spins of $\frac{1}{2}$, each with two states and mirror images of each other. Despite the negative energy issues, he interpreted the two as a matter–antimatter pair [27].

However, there is no bivector in the Dirac equation [28], and introducing one is our departure from the usual development. Multiplying one gamma matrix by the imaginary number, chosen to be $\tilde{\gamma}_s^2 \equiv i\gamma_s^2$, is the only fundamental change introduced in this work. Dirac's gamma field is then replaced by $(\gamma_s^0, \gamma_s^1, \pm\tilde{\gamma}_s^2, \gamma_s^3)$. The subscript $s$ denotes spin spacetime [7], which differs from the Minkowski space, where Dirac's spin is defined. There are two solutions to the complexified Dirac equation: reflective or mirror states. They have no parity and are the matter–antimatter pair Dirac proposed.

Introducing a bivector changes the Clifford algebra from that of the Minkowski space, $\mathbb{C}\ell_{1,3}$ to $\mathbb{C}\ell_{2,2}$ for spin spacetime. Dirac's two spins have SU(2) symmetry. Introducing a bivector changes the symmetry to the quaternion group, $Q_8$.

The bivector complexifies the Dirac field, which was the primary motivation behind Twistor theory [29,30]. The Dirac equation becomes non-Hermitian [7], so spin spacetime is complex, giving helicity states as complex conjugates. The $\pm$ in the gamma matrices above is due to complexification.

As shown elsewhere [7], the non-Hermitian Dirac equation further separates into distinct spaces under parity symmetry. One space is of even parity and describes a 2D disc obeying $\mathbb{C}\ell_{1,2}$. The second space is the $S^3$ hypersphere, generating a unit quaternion that spins the axis of linear momentum, Figure 1. In free flight, a spin is a spinning disc of angular momentum. Despite the coupling of the two fermionic axes to give a composite boson, the helicity causes the polarization to average out. In free flight, an electron is a boson of odd parity, $e_B^-$, with only helicity states of L or R, Figure 1.

The upper central frame of Figure 3 shows a free-flight boson formed from coupling the mirror states of the two fermionic axes (3,1). When encountering a field, illustrated by the longer arrow, the indistinguishable is broken. Then, depending on its orientation relative to the boson spin, either the boson preceeses uncoupled, lower middle panel, or if the field is closer to one of the two fermionic axes, the boson decouples into a fermion, as shown in the left and right panels. This recovers Dirac spin, which obeys Dirac's equation, and is the spin-$\frac{1}{2}$ we measure.

How the resonant spin decouples depends upon the filter strength and its orientation relative to the resonant spin. This is discussed in the following section.

**Spin decoupling**.

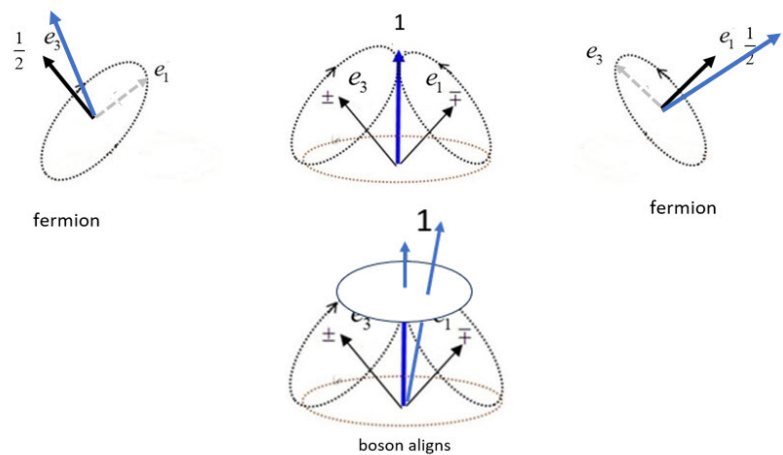

**Figure 3.** The longer arrow denotes the direction of the polarizing field. **Middle top:** The two mirror states in free flight, $e_1$ and $e_3$, couple to give a boson spin-1. **Left and right:** The fermionic axis closer to the field axis aligns, and the boson decouples. These two panels do not indicate two particles but one Q-spin depicted with one or the other axis aligned with the field. **Middle bottom:** When the boson spin is close to the field, it initially precesses as a spin one without decoupling.

*2.2. Q-spin in a Polarizing Field*

The complementary attributes of spin, polarization, and coherence simultaneously exist. We define Q-spin, Σ, possessing both properties,

$$\Sigma = \sigma + \underline{\underline{\mathbf{h}}} \tag{6}$$

There are two components, one for each axis, $\Sigma_k = \Sigma \cdot e_k$, $(k =, 1, 3)$ [7]. Contracting this complex spin with a field vector gives a unit quaternion, $\mathbf{a} \cdot \Sigma$. We combine these two axes, which constructively interfere to produce a pure resonance spin being a composite boson of spin-1,

$$\Sigma_{31} = e_3 \Sigma_3 + e_1 \Sigma_1 \tag{7}$$

The expectation values are obtained for a single Q-spin in a vector field being the product of three quaternions (see [7] Equation (43)),

$$\mathbf{a} \cdot \langle \Sigma_{31} \rangle = \exp\left(i\left(\frac{\pi}{4} + (\theta - \theta_a)\right) Y\right)$$
$$= e^{i\frac{\pi}{4}Y} e^{+i\theta Y} e^{-i\theta_a Y} \tag{8}$$

The first quaternion is a phase (see Equation (30) of [7]); the second is a geometric factor that orients the spin disc in the Body Fixed Frame, BFF; and the last is a field quaternion that sets the field position in the Laboratory Fixed Frame, LFF. Since Q-spin has structure, the BFF and LFF are needed.

There are several practical ways of expressing Equation (8) [7], and these lead to two mechanisms for the transition from a free-flight boson to a measured fermion, as depicted in Figure 3.

Q-spin has three axes of angular momentum: the two fermions, each with a vector magnetic moment of $\mu$, and the one bivector boson, with a magnetic moment of $2\mu$. Introducing a magnetic field causes the vector with the greatest attraction to align and, hence, the others to lie orthogonal to it. This is an example of the classical Least Action Principle used in the simulation: the action is minimized for the strongest attraction. Note that the two fermionic axes have opposite magnetic moments; the field orientation determines whether the state is plus or minus.

*2.3. Separating an EPR Pair*

Initially, in a singlet state at the source, an EPR pair separates, and the two independent particles move towards their respective filters. In the process, they conserve energy, linear momentum, angular momentum, and helicity, see Figure 4. Additionally, correlation is conserved [8] without non-locality. The discs depict the separation process, with the middle disc showing that the two spins are anti-parallel at the source. The angle $\theta$ orients Alice's spin in the LFF and is the same angle it makes on the Bloch sphere. The vector, $e_3$, on the middle disc gives the relationship between the LFF and the BFF. Bob's spin is anti-parallel to Alice's, so his spin is oriented at $\theta \pm \pi$. The azimuthal angle is zero since the spin disc and the polarizers are coplanar.

The left and right discs in Figure 4 depict the two spins after separation. They are beyond the range of any interaction. The common angle $\theta$ means the two BFF frames are identical, except they have opposite handedness. Alice's spin moves right, and Bob's moves left, along the axis of linear momentum, $Y$. This correlation is maintained as the helicity spins both spin axes with identical frequency in the same direction. Alice is in a right-handed frame by the right-hand rule, and Bob is in a left. Complex conjugation relates to the two frames, so they have the same sense of precession, even though their helicities are opposite when viewing Bob's from an RH frame.

**Separation of an EPR pair.**

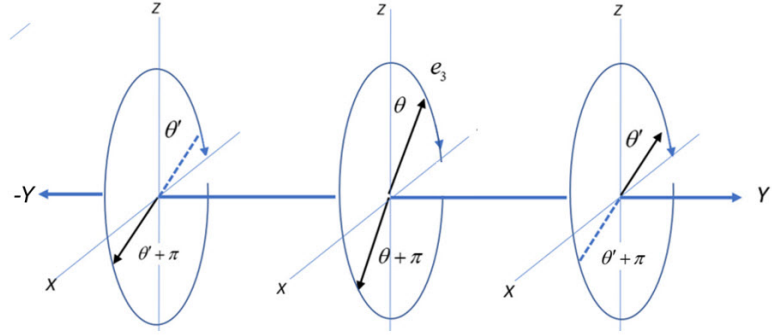

**Figure 4.** Alice goes right in an RH coordinate frame, and Bob goes left in an LH frame. The two spins precess in the same direction, either clockwise or anti-clockwise. Changing Bob's frame from L to R means the helicity of the two particles is opposite. In the center figure, Alice and Bob start with the same but opposite orientation in the BFF indicated by $e_3$.

The frequencies and LFF for both Alice and Bob must be identical for helicity conservation between a separated EPR pair. Otherwise, they would decohere. After spacelike separation, they remain in phase after both have precessions over many periods. Alice and Bob carry the angle $\theta$ and is the only source of correlation between the two. The two spins are produced as anti-parallel and remain so until they encounter their filters.

In the following, we obtain the correlation from coincident experiments in different cases.

*2.4. Correlation from an EPR Pair*

In the absence of the filters **a** and **b**, the correlation is given by the product of Q-spins for Alice and Bob,

$$E(\text{free-flight}) = \frac{1}{2}\left(\langle \Sigma_{31}^A \rangle \langle \Sigma_{31}^B \rangle^* + \langle \Sigma_{31}^A \rangle^* \langle \Sigma_{31}^B \rangle\right)$$
$$= \frac{1}{2}\exp\left(i\left(\frac{\pi}{2}+\theta\right)Y\right)\exp\left(i\left(\frac{\pi}{2}-\theta\right)Y\right) + c.c. = -1 \quad (9)$$

The spins are anti-correlated at separation, which requires $\theta$ to differ between Alice and Bob by $\pi$, see Figure 4. Also, the helicities of Alice and Bob's spins are opposite, expressed by the complex conjugation. The second term reverses the helicity of both, so the correlation is real. This is similar to forming polarized light with photons. As expected, after leaving a common source in free-flight before encountering filters, the correlation between Alice and Bob is $-1$, consistent with the two spins remaining anti-parallel, coherent, and anti-correlated from the source up to the filter.

*2.5. Approaching a Filter*

As the spins of Alice and Bob approach their randomly set filters, the correlation between the two resonance states is

$$E(a,b) = \mathbf{a} \cdot \frac{1}{2}\left(\langle \Sigma_{31}^A \rangle \langle \Sigma_{31}^B \rangle^* + \langle \Sigma_{31}^A \rangle^* \langle \Sigma_{31}^B \rangle\right) \cdot \mathbf{b}$$
$$= \frac{1}{2}\exp\left(i\left(\frac{\pi}{2}-(\theta_a-\theta)\right)Y\right)\exp\left(i\left(\frac{\pi}{2}+(\theta_b-\theta)\right)Y\right) + c.c. \quad (10)$$
$$= -\cos\theta_{ab}$$

This is identical to the correlation from a singlet state that maintains entanglement, $-\cos\theta_{ab}$, but Equation (10) is a product state with no entanglement. Using Q-spin, the correlation is maintained by the common angle $\theta$ without any non-local connectivity between Alice and Bob. Decoherence occurs when $\theta$ for Alice and Bob differ. However, EPR coincidence experiments maintain correlation over space-like separations, from which we conclude the helicity maintains a common frequency between Alice and Bob, and do not decohere, *i.e.*, $\theta$ remains stable in free flight.

However, to get this result, both the spins of Alice and Bob must be complex. If one spin is polarized, there is no helicity, and only the scalar part of the quaternion is present,

$$\exp\left(i\left(\frac{\pi}{2} + (\theta_b - \theta)\right)Y\right) = e^{i\frac{\pi}{2}Y}\exp(i(\theta_b - \theta)Y) \tag{11}$$
$$\xrightarrow{\text{no helicity}} i\cos(\theta_b - \theta)$$

We separated the $\frac{\pi}{2}$ phase needed for anti-correlation. Only the product state survives even if one spin is coherent,

$$\begin{aligned}E(a,b) &= -\frac{1}{2}\exp(i(\theta_a - \theta)Y)\cos(\theta_b - \theta) + c.c.\\ &= -\cos(\theta_a - \theta)\cos(\theta_b - \theta)\end{aligned} \tag{12}$$

To obtain the full correlation, Equation (10), Alice and Bob's particles must be boson spins. If one or the other has decoupled into a fermion, only a product state is possible. However, since the boson has twice the magnetic moment of a fermion, and the two fermions are bound to a boson by internal spin–spin coupling, we surmise the boson resists decoupling until the field is close to a fermion axis and also dominates the spin–spin coupling. For these reasons, the mustache function makes contributions to the correlation over a wider range of $\theta_{ab}$ than the fermion axes. Close to multiples of $\frac{\pi}{2}$, the boson has decoupled. At zero and $\pi$, the correlation is respectively anti-correlated with $-1$ and correlated with $+1$. At $\frac{\pi}{2}$ and $\frac{3\pi}{2}$, the fermions are orthogonal, and there is no correlation.

The correlation from coherence can only be measured when Alice and Bob's spins are oriented within the same 45° wedge as seen in Figure 5. The boson axis, labeled with 1's, represents Alice and Bob's anti-parallel bosons as they approach their (space-like separated) filters. When they respond, the anti-parallel bosons reorient, with both pulled to the nearest field axis where they align and precess. The filtering processes for Alice and Bob are entirely independent and, therefore, local.

**Maximum correlation from CHSH,**

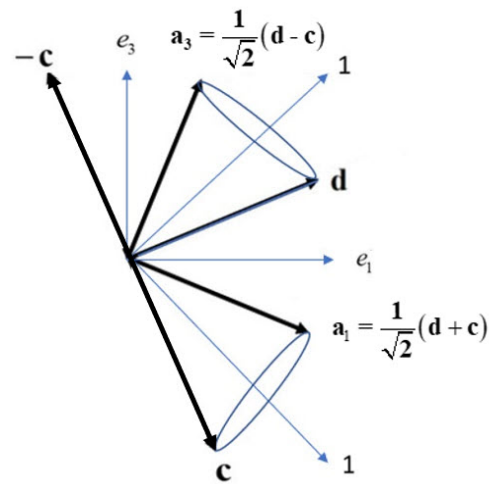

**Figure 5.** Showing Alice and Bob's spins by looking along their common axis of linear momentum, orthogonal to the screen. The heavy lines are possible filter settings in the LFF, XZ plane. Bob's settings are $-\mathbf{c}$, $\mathbf{d}$ or $\mathbf{c}$. Alice has settings of either $\mathbf{a_3}$ or $\mathbf{a_1}$. Also shown, bisecting two of the four 45° cones, are arrows labeled "1". These are the two antiparallel boson spins of Alice and Bob impinging on their filters. Over spacetime, there is little decoherence, so within the two 45° wedges, Alice and Bob's spins can both be bosons.

*2.6. Interpretation of the CHSH Inequality*

Coupling of angular momentum is standard in all spectroscopes [31], including transport properties of gases [32,33]. In our case, the resonance spin is stable in free-flight but will eventually decouple as it approaches a polarizing field. At a given distance, the decoupling depends upon the orientation of a spin relative to the filter and its strength relative to the spin–spin coupling within Q-spin.

Consider now that settings for both Alice and Bob are shown in Figure 5. Bob's is one of $-\mathbf{c}, \mathbf{d}$, and $\mathbf{c}$, and Alice's settings is one of either $\mathbf{a_3}$ or $\mathbf{a_1}$, with $\mathbf{a_3} = \frac{1}{\sqrt{2}}(\mathbf{d} - \mathbf{c})$ and $\mathbf{a_1} = \frac{1}{\sqrt{2}}(\mathbf{d} + \mathbf{c})$. There are four experiments expressed in the CHSH inequality, whence Alice and Bob have correlations of $E(-\mathbf{c}, \mathbf{a_3})$, $E(\mathbf{d}, \mathbf{a_3})$, $E(\mathbf{d}, \mathbf{a_1})$, and $E(\mathbf{c}, \mathbf{a_1})$. Q-spins impinge on the filters at random angles of $\theta$ from a random source. The figure shows two possible orientations of the anti-correlated bosons, each labeled by a "1", which bisect two of the filter cones. We chose these settings because Alice and Bob's bosons are then close to their filter, only 22.5° away, and remain bosons as they align with their filter and precess about it's axis. Therefore, their contribution to the correlation is from coherence, coh. If, in contrast, an experiment is, say, $E(\mathbf{c}, \mathbf{a_3})$, then Alice is still only 22.5° from $\mathbf{a_3}$, but now Bob is 112.5° from his. Whereas Alice remains a boson, Bob's boson must decouple when encountering his filter at $\mathbf{c}$. Only polarization, pol, then contributes to the correlation, Equation (12).

When considering the possible outcomes from coincidences, there are four types: pol:pol, pol:coh, coh:pol, and coh:coh. Only the coh:coh coincidences contribute to the correlation from coherence.

The maximum correlation for the *S*=CHSH inequality for the four experiments is determined by the filter setting as shown in Figure 5, giving,

$$\begin{aligned} s &= \mathbf{a_1} \cdot (\mathbf{d} + \mathbf{c}) + \mathbf{a_3} \cdot (\mathbf{d} - \mathbf{c}) \\ &= \tfrac{1}{\sqrt{2}}(\mathbf{d} + \mathbf{c}) \cdot (\mathbf{d} + \mathbf{c}) + \tfrac{1}{\sqrt{2}}(\mathbf{d} - \mathbf{c}) \cdot (\mathbf{d} - \mathbf{c}) \\ &= 2\sqrt{2} \end{aligned} \quad (13)$$

Consider the correlation from one experiment, $E(\mathbf{d}, \mathbf{a_3})$ shown in Figure 5. If Alice and Bob's spins reorient as bosons, their precessions will look as shown in Figure 6. Spin makes an angle of $\cos \chi = \frac{m}{\sqrt{S(S+1)}}$ with a field direction. That is $\chi = 45°$ for spin-1 bosons, as shown in Figure 6. A spin $\frac{1}{2}$ has $\chi = 54.74°$, supporting the assertion that the bosons remain intact within the cone.

**Strong correlation between Alice and Bob.**

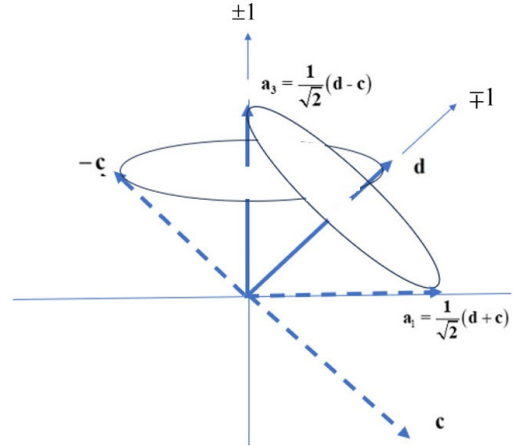

**Figure 6.** Displaying the maximum correlation from coherence. Alice sets her filter to $\mathbf{a_3}$, and her boson spin of 1 aligns. Bob has two settings that will give the maximum violation by applying his filter either along $\mathbf{d}$ or along $-\mathbf{c}$ at $45°$ from $\mathbf{a_3}$. Alternately, Alice can set her filter angle to $\mathbf{a_1}$, and Bob has two filter settings, $\mathbf{d}$ and $\mathbf{c}$ that lead to the maximum correlation. Note that a spin of magnitude 1 precesses about its axis with an angle of $45°$.

In Figure 5, the filter angles for Alice and Bob are those that give the maximum correlation. Any three filter angles for Bob and any two for Alice can also be used, but as they move from those in Figure 5, the CHSH correlation will decrease.

### 3. Simulation Model

In contrast to a structureless point particle, Q-spin displays four internal motions with four axes. First, the axis of linear momentum is spun by a quaternion. The two fermion axes can precess about their axes, which maintain mirror symmetry, and couple to give the fourth axis: the boson.

In [7], Q-spin in a polarizing field, Equation (8), is expressed in different ways which show the projections of Alice and Bob's spins onto the boson axis, $e_{13}$, and the decoupled axes, $e_3$ and $e_1$. From these expressions in the LFF, the correlation between Alice and Bob is due to a competition between axes using the Least Action Principle. Consider Equation (14) of [7] given by,

$$\mathbf{a} \cdot \langle \Sigma_{31} \rangle = \frac{1}{\sqrt{2}} \left( \cos(\theta_a - \theta) \exp\left(+i\frac{\pi}{4}Y\right) + \sin(\theta_a - \theta) \exp\left(-i\frac{\pi}{4}Y\right) \right) \quad (14)$$

The two coefficients are projections of the two BFF spin axes along the field axis,

$$\begin{aligned} \mathbf{a} \cdot e_3 &= \cos(\theta_a - \theta) \\ \mathbf{a} \cdot e_1 &= \sin(\theta_a - \theta) \end{aligned} \quad (15)$$

and the larger magnitude will align.

#### 3.1. Quaternion Algorithm

Consider a boson in a polarizing field (upper middle panel of Figure 3) showing the two coupled axes giving the boson. It has two axes since it comes from a bivector. We can deterministically know which axis will align when it decouples. Before decoupling, we must double the angle to account for the double Larmor frequency and write the two coupled boson axes as,

$$2\operatorname{Re}\left(\mathbf{a} \cdot \left\langle \Sigma_{31}^1 \right\rangle\right) = \overbrace{\cos(2\theta_a - \theta)}^{Z} + \overbrace{\sin(2\theta_a - \theta)}^{X} \quad (16)$$

The larger axis will become polarized, and its sign gives a $+$ or $-$ click. Without doubling the angle, $\theta_a$, the simulation fails. The doubling gives the mustache function two periods over one period of $\theta_{ab}$ in Figure 7. The CHSH correlation for the full mustache is 1.

Of course, Bob's coincidences are found the same way and combine with Alice's into equal coincidences $(\pm, \pm)$ or unequal coincidences, $(\pm, \mp)$.

If Alice and Bob's spins are outside the 45° wedge, the boson spin will start to decouple. When it lies 45° from its field, it completely decouples, its orthogonal partner averages away, and only one axis remains, Figure 3, left and right panels. The usual Larmor frequency for a fermion is proportional to $\mu$ and does not double the angle. Then, we determine if that axis is positive or negative, indicating a plus or minus click.

Using a vector gives the triangle, Figure 7, with a CHSH = 2. Using a bivector gives the mustache function with CHSH = 1.

To summarize, coherence depends upon the two axes of the bivector. First, determine the larger axis,

$$|\text{axis (X)}| \overset{?}{<>} |\text{axis (Z)}| \tag{17}$$

and then use the sign of that axis to determine the click values. Since $\theta$ varies from 0 to $2\pi$, the magnitudes and signs of the projected axes vary.

For polarization, use only one axis, and we choose, e.g.,

$$\cos(\theta_a - \theta) \overset{?}{<>} 0 \tag{18}$$

The sign, again, deterministically gives spin up or down.

**Full correlation**

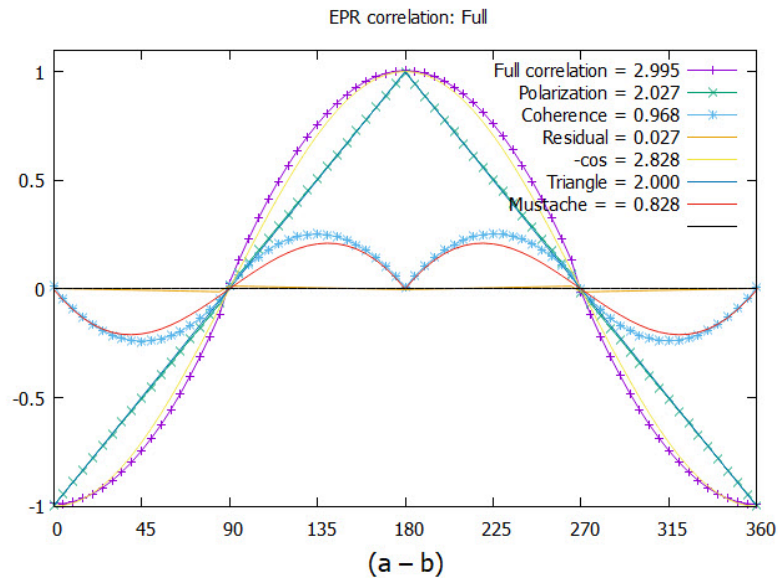

**Figure 7.** Plotting EPR correlation versus the angle difference $(\theta_a - \theta_b)$. The points give the results of the simulation. The CHSH values are listed and the full correlation is the sum of polarization, (the triangle) and the coherence, (the mustache). Note the hardly discernible residual polarization correlation along the horizontal axis, which shows the contribution extracted from the polarization, $2.027 - 2.000 = 0.027$.

The simulation determines the correlation between classical axes. It is not a quantum simulation; instead, it is a model that the treatment here suggests. It is known that classical simulations can account for the correlation without non-locality [34–38]. We note that the total correlation from polarization and coherence gives a total of CHSH = 3, in contrast to that from an entangled state of $2\sqrt{2}$. As discussed in [20], this violation of the the Tsirel'son bound, [39] is due to the entangled singlet state being an approximation.

*3.2. Simulation Results*

There are a number of simulations which study EPR correlations, providing insight to hidden variables, locality and contextuallity, [40–44]. The FORTRAN and C code, available with this paper, determines clicks from polarized states and coherent events using the above algorithms. Additionally, we provide several text files containing the simulation's outputs, along with the plotting code for Gnuplot.

The results are given in Figure 7. The simulated correlation from either polarization axis gives the polarization as the triangle, with a CHSH = 2.027. The mustache coherence, which depends on the coupling of the two axes into one axis, is shown in the same figure and gives a CHSH = 0.968. The simulation gives a total CHSH value of 2.995, or almost 3. We find more correlation than from a singlet state, CHSH = 2.828.

The theoretical CHSH value for polarizations is 2, whereas in Figure 7, it is 2.027. As shown in Figure 8, residual correlation remains. A cross-over occurs close to the horizontal axis. It is hardly discernible but is a natural feature in the model. Its sign is in tandem with both the polarizations and the coherences. We assert that this residual contribution is due to the correlation from the polarization states. If confirmed, the classical polarization exceeds BI [9] by a small amount and violates the theorem again [11].

We expected the cross-over in Figure 8 to be at zero and have not determined why it is shifted lower.

**Residual quaternion correlation**

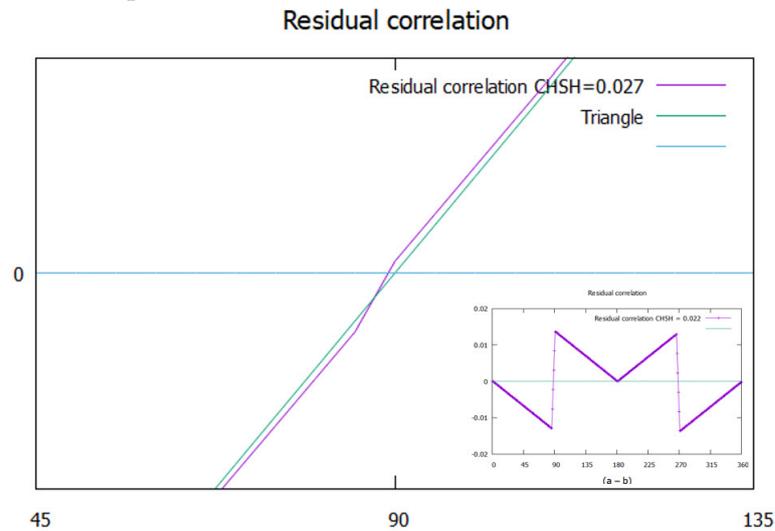

**Figure 8.** A blow-up of the simulated polarization, with the cross-over, compared to an exact triangle, the straight line. The insert shows the difference between the triangle and the cross-over over $2\pi$.

*3.3. Determining the Correlation*

Polarization and coherence are complementary properties, meaning they are not manifest simultaneously [45,46]. Pauli stated [47], "Intuitively, observables are complementary if the experimental arrangements allowing their unambiguous definitions are mutually exclusive". Coincidence EPR experiments are not yet able to distinguish polarization events from coherence events. In our case, the angle, $\theta$, is taken as random but can be in the experimenter's control and used, in principle, to distinguish polarization from coherence. In this sense, Pauli's statement is consistent here.

The correlation between equal coincidence events $(\pm, \pm)$, $N_{eq}$, and unequal events $((\pm, \mp))$, $N_{nq}$ is given by

$$E(a,b) = \frac{N_{eq} - N_{nq}}{N_{tot}} \tag{19}$$

which gives the minus cosine similarity in EPR coincidence experiments.

Using Q-spin, the coincidence events are complementary. Sometimes, an event gives a boson correlation; other times, it gives a polarization correlation. However, complementary properties do not exist in the same space. Polarization belongs to a convex set, a disc with $\mathbb{C}\ell_{1,2}$. Quaternion space, the $S^3$ hypersphere, is also convex. Polarization and coherent states are pure since we treat a single EPR pair. The two complementary spin attributes are extreme points in their distinct convex spaces.

With no mixed states, we must only distinguish between polarized (pol) and coherent (coh) states. We introduce Boolean operators that ensure complementarity,

$$\left[\delta_p, \delta_c\right] \quad (20)$$

where $\delta_p = 1, \delta_c = 0$ or $\delta_p = 0, \delta_c = 1$, with the sum of the contributions being the total number of events, $N_{tot}$.

Suppose that it is possible to filter the system such that the experiments can distinguish between the two, then we can write the total correlation as collecting the different events in different bins,

$$\begin{aligned} E(a,b) &= \frac{\delta_p N_{eq}^p + \delta_c N_{eq}^c - \delta_p N_{nq}^p - \delta_c N_{nq}^c}{N_{tot}} \\ &= \frac{N_{eq}^p - N_{nq}^p}{N_{tot}} \delta_p + \frac{N_{eq}^c - N_{nq}^c}{N_{tot}} \delta_c \\ &= E_p(a,b)\delta_p + E_c(a,b)\delta_c \end{aligned} \quad (21)$$

The polarization and coherence correlation are evaluated and added as shown in Figure 7.

Note that the value of a correlation is independent of the total number of clicks, so for the calculation, we did the following: First, in Equation (21), assume $\delta_p = 1$ and $\delta_c = 0$, and the polarization algorithm gives the triangle in Figure 7. Then we set $\delta_p = 0$ and $\delta_c = 1$, and the coherence algorithm gives the mustache function.

Correlation from an entangled singlet is divided between correlation from polarization and coherence. This is consistent with the Conservation of Geometric Correlation [8], which states that the total correlation at the source is conserved between particles upon local separation.

The experimental challenge is distinguishing between polarization and coherence for a single particle and controlling the boson decoupling mechanism. In principle, at a source, its orientation is determined by $\theta$. By definition, a single particle is in a pure state and remains so after leaving the source until encountering a field. The filter angle, $\theta_a$, can then be chosen. Finally, bosons have double the Larmor frequency of the fermions as a possible handle. A goal might be to find methods that would consistently and deterministically produce outcomes of $|+,\hat{\mathbf{n}}\rangle$ or $|-,\hat{\mathbf{n}}\rangle$. Spin is a stable property of many particles and its helicity does not decohere over spacetime.

### 3.4. The Mustache Function

The correlation between bosons is responsible for the mustache function. The coherent correlation, $E_q$, resembles two opposing sine waves reflected at $\pi$. The following gives a good fit,

$$E_q = \begin{cases} -\frac{1}{4}\sin(2\theta_{ab}) & 0 \le \theta_{ab} \le \pi \\ +\frac{1}{4}\sin(2\theta_{ab}) & \pi < \theta_{ab} \le 2\pi \end{cases} \quad (22)$$

This is plotted in Figure 9 along with the simulated data points, which match Equation (22). Also plotted for comparison is the correlation from quantum theory. The simulation gives more correlation, CHSH = 1, than obtained from the singlet state and QM, CHSH = 0.828, the difference, as we mentioned, is due to the singlet state being an approximation [20].

The doubling of the filter angles, $2\theta_{ab}$, is due to the boson magnetic moment of $2\mu$, which gives two periods over $\theta_{ab}$. This doubling gives credence to the presence of a boson spin-1, with a magnetic moment of $2\mu$, being responsible for the mustache, and not a fermion, spin-$\frac{1}{2}$, with the magnetic moment of $\mu$. In the simulation, anti-correlation means the spin vectors for Alice and Bob have opposite signs, Equation (18), and this means the magnetic quantum numbers, *m*, for Alice and Bob are opposite. The *m* values are expressed by the sign of an axis in a magnetic field. The first period in Figure 7 starts at $\theta_{ab} = 0$ where Alice and Bob's spins are anti-correlated. As $\theta_{ab}$ moves to $\frac{\pi}{4}$, the anti-correlation means the coherence correlation is negative. At that point, the two bosons begin to correlate, rendering the boson correlation positive.

The second period starts at $\pi$, but the bosons remain correlated up to $\frac{3\pi}{2}$, after which they are again anti-correlated. Therefore, the boson correlation in the second period starts off positive. The sign of the boson correlation is in tandem with the sign of polarization correlation, *i.e.*, the sign of the triangle. We note the residual correlation of 0.027, Figure 8 has the same sign changes. Each sine curve gives the correlation between the bosons of Alice and Bob. This maximizes at multiples of $\frac{\pi}{4}$ where both are bosons.

We view the correlations from the polarization (pol) and coherence (coh) as distinct. By dropping the coherent terms, only fermion electrons remain, and the correlation is the triangle in Figure 7. Putting back the boson spins gives the mustache function, Equation (22). However, recall that to obtain correlation from coherence, Alice and Bob must be bosons, and both in coherent states, coh-coh. There are four types of coincidences between them: pol-pol, pol-coh, coh-pol, and coh-coh. Therefore, only $\frac{1}{4}$ of the coincidences lead to coherence, which is the origin of that factor in Equation (22).

**Comparison of the Mustache function with $\sin(2\theta)$**

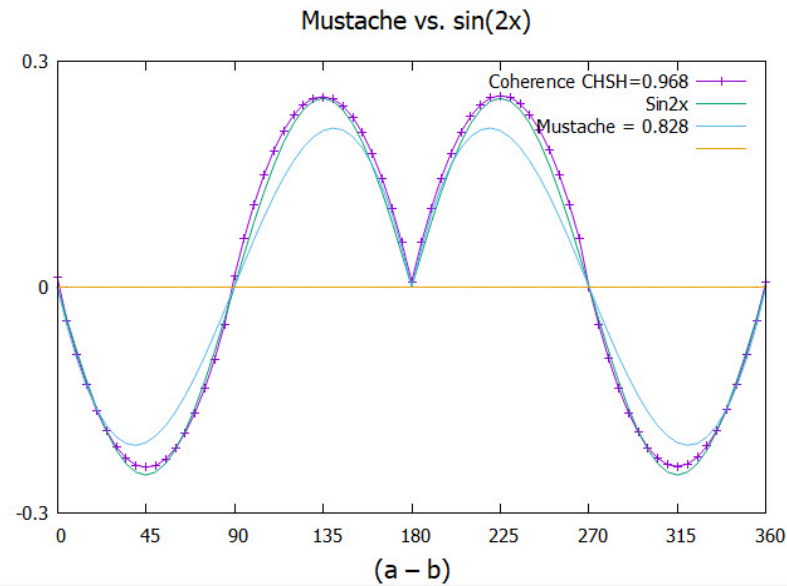

**Figure 9.** The $E_q$ function given in Equation (22) compared to the simulation. Also shown is the smaller correlation from quantum theory.

## 4. Discussion

We find that the two complementary properties exist simultaneously. This supports Einstein in the famous Einstein–Bohr debates [22]. However, only one property can be realized simultaneously, supporting Bohr's notion of complementarity [48]. Coherence and polarization are incompatible elements of reality. They influence each other. The spinor spins the polarization; the indistinguishable two polarization axes create the mirror states for the helicity to know which way to spin.

Figure 10 lists four values of the CHSH inequalities. The value of 2 quantifies the Infamous Boundary [49] between polarization and coherence. Quantum theory gives the value of $2\sqrt{2}$, which is less than that from the simulation of 2.995, which is close to 3. The value of 3 results from modeling the three axes of Q-spin, suggesting a CHSH = 1 for each axis.

The CHSH values leading to 2.995 from the simulation are $E\left(\frac{\pi}{4}\right) = -0.746$ and $E\left(\frac{3\pi}{4}\right) = +0.758$. This suggests a convergence to $\frac{3}{4}$, which is $\frac{1}{4}$ correlation per axis. This leads to a CHSH value of 3, which supports equal contributions from each axis.

**Some CHSH values**

| Classical | Quantum theory | Quaternion spin | Mother Nature |
|---|---|---|---|
| 2.000 | 2.828 | 2.995 | 3 |

**Figure 10.** Values of the CHSH inequalities from classical to Nature.

*4.1. Spin*

In this treatment, helicity replaces non-local entanglement. Q-spin replaces Dirac spin. Both these features come from changing the symmetry from SU(2) to $Q_8$. The visualization of Q-spin, which is geometrically identical to a photon in free flight, see Figure 1, carries a 2D spin plane that forms its own "World Sheet" [50]. The 2D structure of Q-spin admits anyons [51], consistent with a boson and a fermion on one particle. This is not possible in 3D or for a point particle. However, usually, anyons are depicted as the braiding caused by the interchange of two identical particles. Here, the braiding is caused by exchanging two identical axes on the same particle.

When the Dirac equation is changed to include Q-spin, the two resulting mirror states have no parity, only reflection, but they combine into states of even parity polarization and odd parity coherence. The mirror states that Dirac interpreted as a matter-antimatter pair represent, for Q-spin, two magnetic axes on one particle. The two reflective axes are orthogonal, in phase, and precess oppositely with equal energy as shown in the upper middle panel in Figure 3. Q-spin resolves Dirac's negative energy issue.

If Q-spin was formed at the Big Bang, producing no anti-matter, then the baryon asymmetry problem [52] is resolved.

Spin carries helicity as an element of reality, and quaternions exist in the $S^3$ hypersphere of four spatial dimensions. Spin extends to these spaces beyond our visualization but remains an element of reality beyond our dimension. The only part of a quaternion visible to us is the stereographic projection from $S^3$ onto our Minkowski space. From that, we observe a spinning axis, giving our usual spin vector, **S**.

According to our treatment, EPR is correct in asserting that quantum theory is incomplete. Quantum mechanics is a theory of measurement but not of Nature. It does not include those higher dimensions of the $S^3$ hypersphere needed for the coherence; it does not include anti-Hermitian operators as elements of reality, nor the complexity of spin spacetime. Despite being contrary to accepted philosophy, the treatment here gives an alternate, deterministic, and locally realistic description of the microscopic.

In calculating expectation values, the state operator includes only the polarization, and not the bivector for helicity. This is justified because a bivector is not physically observable in our spacetime.

In free flight, Q-spin acts like a boson with odd parity, and in a perturbing field, it acts like a fermion with even parity.

*4.2. Bell*

Bell's theorem states that no local theory can exceed his inequality and asserts in his own words [11],

> If [a hidden-variable theory] is local, it will not agree with quantum mechanics; if it agrees with quantum mechanics, it will not be local.

Since we have shown that coherence accounts for the violation of BI, thereby obviating non-locality, this statement by Bell cannot be supported. Q-spin disproves Bell's Theorem by counterexample. Bell's Inequalities are classical and quantitatively express correlations between real variables in our 3D space. His classical theorem proves that a classical system cannot violate his upper bound. In contrast, Q-spin is quantum, with two complementary attributes in different convex spaces. It carries helicity, or coherence, in addition to polarization, and the variables are complex. Bell's theorem does not apply to quantum systems, and no conclusions concerning non-locality and quantum theory can be drawn from any of his works [11]. Only his classical inequalities are relevant to QM, because they quantitatively define the "Infamous boundary" [49] between the classical and quantum worlds via a CHSH value of two.

In this work, there are no Hidden Variables. The only variable is local, $\theta$, relating spin and Minkowski spaces. When BI is violated [53], non-locality is not the cause. Instead, the opposite follows, as shown here: the violation is evidence of local realism.

However, the simulated CHSH gives a minor violation, whereby 2.027 exceeds BI for polarization, Figure 8. This shows that a local realistic theory can violate BI [34–38].

## 5. Conclusions

The existence of Q-spin requires accepting that Nature has a reality beyond our spacetime, that there is information lost [54], and that there are properties in Nature we cannot observe. Some limitations of quantum mechanics can be addressed by replacing Dirac spin with Q-spin and entanglement with helicity.

However, we can show that an entangled singlet state is an approximation [20]. For this reason, the quantum correlation of $2\sqrt{2}$ is less than the simulated value of CHSH = 3. This violates Tsirel'son's bound [39]. Entanglement is a fundamental property of quantum theory but not of Nature. It simplifies calculations by dropping coherent terms, but the price is reduced correlation and lost structure.

Q-spin, if accepted, changes our fundamental view of Nature and requires a re-examination of areas of quantum theory that have hitherto relied on Bell's Theorem to justify non-local connectivity, *e.g.* [1,4], [14,17] and [55–57].

The violation of Bell's inequalities is not evidence for non-locality, an impossible concept to grasp, but rather evidence for the existence of helicity and other properties of Nature that constitute local realism.

**Supplementary Materials:** The following supporting information can be downloaded at: https://www.mdpi.com/article/10.3390/quantum1010000/s1. It contains the simulation programs C and FORTRAN. Additionally, Gnuplot code is given along with the .text files that give most of the figures in this paper.

**Funding:** This research received no external funding.

**Data Availability Statement:** No new data were created or analyzed in this study. Data sharing is not applicable to this article.

**Acknowledgments:** The author is grateful to Pierre Leroy (programmer) and Chantal Roth (programmer) for their help and patience with simulation methods. I thank Pierre, who converted the FORTRAN Program to C.

**Conflicts of Interest:** The author declares no conflicts of interest in writing the manuscript or in the decision to publish the results.

## References

1. Clauser, J.F.; Horne, M.A.; Shimony, A.; Holt, R.A. Proposed experiment to test local hidden-variable theories. *Phys. Rev. Lett.* **1969**, *23*, 880.
2. Aspect, A.; Jean, D.; Gérard, R. Experimental test of Bell's inequalities using time-varying analyzers. *Phys. Rev. Lett.* **1982**, *49*, 1804.
3. Aspect, A. Proposed experiment to test the non separability of quantum mechanics. *Phys. Rev. D* **1976**, *14*, 1944–1951
4. Weihs, G.; Jennewein, T.; Simon, C.; Weinfurter, H.; Zeilinger, A. Violation of Bell's inequality under strict Einstein locality conditions. *Phys. Rev. Lett.* **1998**, *81*, 5039.
5. Einstein, A., Boris P.; Nathan R. Can quantum-mechanical description of physical reality be considered complete? *Phys. Rev.* **1935**, *47*, 777.
6. Greenberger, D.M.; Horne, M.A.; Shimony, A.; Zeilinger, A. Bell's theorem without inequalities. *Am. J. Phys.* **1990**, *58*, 1131–1143.

7. Sanctuary, B. Quaternion Spin. Mathematics 2024, 12, 1962. https://doi.org/10.3390/math12131962
8. Sanctuary, B. Spin helicity. *Preprints* **2023**, 2023010571. https://doi.org/10.20944/preprints202301.0571.v3.
9. Bell, J.S. On the Einstein Podolsky Rosen paradox. *Phys. Phys. Fiz.* **1964**, *1*, 195.
10. Dirac, P.A.M. The quantum theory of the electron. *Proc. R. Soc. Lond. Ser. A Contain. Pap. Math. Phys. Character* **1928**, *117*, 610–624.
11. Bell, J.S. *Speakable and Unspeakable in Quantum Mechanics: Collected Papers on Quantum Philosophy*; Cambridge University Press: Cambridge, UK, 2004; pp. 139/147.
12. Fisher, A.D. *A Few Insights into Quantum Entanglement*; 2023. Available online: https://www.researchgate.net/publication/370492920_A_Few_Insights_into_Quantum_Entanglement (accessed on 6 August 2024). https://doi.org/10.13140/RG.2.2.16605.900811.
13. Erhard, M.; Krenn, M.; Zeilinger, A. Advances in high-dimensional quantum entangle-ment. *Nat. Rev. Phys.* **2020**, *2*, 365–381.
14. Bennett, C.H.; Brassard, G.; Crépeau, C.; Jozsa, R.; Peres, A.; Wootters, W.K. Teleporting an unknown quantum state via dual classical and Einstein-Podolsky-Rosen channels. *Phys. Rev. Lett.* **1993**, *70*, 1895
15. Rozenman, G.G.; Kundu, N.K.; Liu, R.; Zhang, L.; Maslennikov, A.; Reches, Y.; Youm, H.Y. The quantum internet: A synergy of quantum information technologies and 6G networks. *IET Quantum Commun.* **2023**, *4*, 147–166.
16. Bennett, C.H.; Brassard, G. Quantum cryptography: Public key distribution and coin tossing. *Theor. Comput. Sci.* **2014**, *560*, 7–11.
17. Korzh, B.; Lim, C.C.W.; Houlmann, R.; Gisin, N.; Li, M.J.; Nolan, D.; Sanguinetti, B.; Thew, R.; Zbinden, H. Provably secure and practical quantum key distribution over 307 km of optical fibre. *Nat. Photonics* **2015**, *9*, 163–168.
18. Ashida, Y.; Gong, Z.; Ueda, M. Non-hermitian physics. *Adv. Phys.* **2020**, *69*, 249–435.
19. Wang, C.; Fu, Z.; Mao, W.; Qie, J.; Stone, A.D.; Yang, L. Non-Hermitian optics and pho-tonics: From classical to quantum. *Adv. Opt. Photonics* **2023**, *15*, 442–523.
20. Sanctuary, B. Quaternion-Spin and Some Consequences. *Preprints* **2023**, 2023121277. https://doi.org/10.20944/preprints202312.1277.v1.
21. Bohm, D. *Quantum Theory*; Prentice-Hall: Englewood Cliffs, NJ, USA, 1951; p. 29, Chapter 5 Section 3, and Chapter 22 Section 19.
22. Jammer, M. *Philosophy of Quantum Mechanics. The Interpretations of Quantum Mechanics from a Historical Perspective*; John Wiley and Sons: New York, 1974.
23. Turnbull, H.W. (Ed.) *Newton to Bentley, February 25 1692/3, The Correspondence of Isaac Newton*; Cambridge University Press: Cambridge, UK, 1961.
24. Einstein, A. *Born-Einstein Letters 1916–1955*: Friendship, Politics and Physics in Uncertain Times. Palgrave Macmillan: Basingstoke and New York, 2014.
25. Mullin, W.J. *Quantum Weirdness*; Oxford University Press: Oxford, UK, 2017.
26. Doran, C.; Lasenby, J. *Geometric Algebra for Physicists*; Cambridge University Press: Cambridge, UK, 2003.
27. Dirac, P.A.M. A Theory of Electrons and Protons. *Proc. R. Soc. Lond. A* **1930**, *126*, 360–365.
28. Peskin, M.E.; Schroeder, D.V. *An Introduction to Quantum Field Theory (Frontiers in Physics)*; Westview Press: Boulder, CO, USA, 1995.
29. Penrose, R. Twistor algebra. *J. Math. Phys.* **1967**, *8*, 345–366.
30. Penrose, R. Solutions of the Zero-Rest-Mass Equations. *J. Math. Phys.* **1969**, *10*, 38–39.
31. Herzberg, G. *Molecular Spectra and Molecular Structure*; Read Books Ltd.: Redditch, UK, 2013; Volume 1.
32. Turfa, A.F.; Connor, J.N.L.; Thijsse, B.J.; Beenakker, J.J.M. A classical dynamics study of Senftleben-Beenakker effects in nitrogen gas. *Phys. A Stat. Mech. Its Appl.* **1985**, *129*, 439–454.
33. Sanctuary, B.C.; Beenakker, J.J.M.; Coope, J.A.R. Influence of nuclear spin cou-plings on the thermomagnetic torque in HD. *J. Chem. Phys.* **1974**, *60*, 3352–3353.
34. Annila, A. (2021). *Quantum Entanglement: Bell's Inequality Trivially Violated Also Classically.* Available at SSRN 4388586.;
35. Annila, A.; Wikström, M. Quantum entanglement and classical correlation have the same form. *Eur. Phys. J. Plus* **2024**, *139*, 1–7.
36. Geurdes, H. Bell's Theorem and Einstein's Worry about Quantum Mechanics. *J. Quantum Inf. Sci.* **2023**, *13*, 131–137. https://doi.org/10.4236/jqis.2023.133007. -
37. Dieks, D. Inequalities that test locality in quantum mechanics. *Phys. Rev. A* **2002**, *66*, 062104.
38. Jakumeit, J.; Hess, K. Breaking a Combinatorial Symmetry Resolves the Paradox of Einstein-Podolsky-Rosen and Bell. *Symmetry* **2024**, *16*, 255.
39. Tsirel'son, B. S. (1987). Quantum analogues of the Bell inequalities. The case of two spatially separated domains. Journal of Soviet Mathematics, 36, 557–570.
40. Werner, R. F. Quantum states with Einstein-Podolsky-Rosen correlations admitting a hidden-variable model. *Phys. Rev. A* **1989**, *40*, 4277–4281.
41. Cabello, A. Proposed experiment to test the foundations of physics. *Phys. Rev. A* **2005**, *72*, 052112.
42. Summers, S. J., and Werner, R. (1987). Bell's inequalities and quantum field theory I. General setting. Journal of Mathematical Physics, 28(10), 2440-2447.
43. Leggett, A. J. (2003). Nonlocal Hidden-Variable Theories and Quantum Mechanics: An Incompatibility Theorem. Foundations of Physics, 33(10), 1469-1493.
44. Valentini, A. (2002). Signal-locality, uncertainty, and the sub-quantum H-theorem. Physics Letters A, 297(5-6), 273-278.
45. Kiukas, J.; Lahti, P.; Pellonpää, J.P.; Ylinen, K. Complementary Observables in Quantum Mechanics. *Found. Phys.* **2019**, *49*, 506–531. https://doi.org/10.1007/s10701-019-00261-3.
46. Busch, P.; Grabowski, M.; Lahti, P. *Operational Quantum Physics, LNP 31*; Springer: New York, NY, USA, 1994;

47. Dirac, P.A.M. *The Principles of Quantum Mechanics. No. 27*; Oxford University Press: Oxford, UK, 1981.
48. Bohr, N. Can quantum-mechanical description of physical reality be considered complete? *Phys. Rev.* **1935**, *48*, 696.
49. Wick, D. *The Infamous Boundary: Seven Decades of Heresy in Quantum Physics*; Springer Science and Business Media: Berlin/Heidelberg, Germany, 2012.
50. Maldacena, J.; Susskind, L. Cool horizons for entangled black holes. *Fortschritte der Physik* **2013**, *61*, 781–811.
51. Wilczek, F. Quantum mechanics of fractional-spin particles. *Phys. Rev. Lett.* **1982**, *49*, 957.
52. Baryon Asymmetry. *Wikipedia*. 2024. Available online: https://en.wikipedia.org/wiki/Baryonasymmetry (accessed on 10 August 2024).
53. Abellán, C., Acín, A., Alarcón, A., Alibart, O., Andersen, C. K., Andreoli, F.; Beckert, A.; Beduini, F.A.; Bendersky, A.; Bentivegna, M.; et al. Challenging local realism with human choices. *arXiv* **2018**, arXiv:1805.04431.
54. Braunstein, S.L.; Arun K.P. Quantum information cannot be completely hidden in correlations: Implications for the black-hole information paradox. *Phys. Rev. Lett.* **2007**, *98*, 080502.
55. Kim, Y.H.; Yu, R.; Kulik, S.P.; Shih, Y.; Scully, M.O. (). Delayed "choice" quantum eraser. *Phys. Rev. Lett.* **2000**, *84*, 1.
56. Gisin, N.; Ribordy, G.; Tittel, W.; Zbinden, H. Quantum cryptography. *Rev. Mod. Phys.* **2002**, *74*, 145.
57. Bashar, M. A., Chowdhury, M. A., Islam, R., Rahman, M. S.; Das, S. K. A review and prospects of quantum teleportation. In Proceedings of the 2009 International Conference on Computer and Automation Engineering, Bangkok, Thailand, 8–10 March 2009; pp. 213–217. https://doi.org/10.1109/ICCAE.2009.77.